\documentclass[conference]{IEEEtran}
\usepackage{blindtext}
\usepackage{graphicx}
\graphicspath{ {images/} }
\usepackage{multirow}
\usepackage{bigstrut}
\usepackage[caption=false,font=normalsize,labelfont=sf,textfont=sf]{subfig}
\usepackage{xcolor}
\usepackage{soul}

\ifCLASSINFOpdf
  \DeclareGraphicsExtensions{.pdf,.jpeg,.png}
\else
\fi

\usepackage[cmex10]{amsmath}
\usepackage{booktabs}

\usepackage{cite}

\ifCLASSINFOpdf
\else
\fi
\IEEEoverridecommandlockouts

\usepackage{tikz}
\usepackage{textcomp}
\usepackage{hyperref}
\usepackage{lipsum}



\makeatletter
\newcommand*\titleheader[1]{\gdef\@titleheader{#1}}
\AtBeginDocument{%
  \let\st@red@title\@title%
  \def\@title{%
    \bgroup\normalfont\large\centering\@titleheader\par\egroup
    \vskip1.5em\st@red@title}
}
\makeatother

\newcommand\copyrighttext{%
  \footnotesize \textcopyright 2018 IEEE. Personal use of this material is permitted. Permission from IEEE must be obtained for all other uses, in any current or future media, including reprinting/republishing this material for advertising or promotional purposes, creating new collective works, for resale or redistribution to servers or lists, or reuse of any copyrighted component of this work in other works.
 DOI: \href{<http://tex.stackexchange.com>}{<DOI No.>}}
\newcommand\copyrightnotice{%
\begin{tikzpicture}[remember picture,overlay]
\node[anchor=south,yshift=10pt] at (current page.south) {\fbox{\parbox{\dimexpr\textwidth-\fboxsep-\fboxrule\relax}{\copyrighttext}}};
\end{tikzpicture}%
}

\usepackage{url}

\usepackage{devanagari}
\usepackage{textcomp}

\title{MASTISK : \dn mE-t\309wk }
\titleheader{Reduced pre-print version. Full extended version of this work will be published in the proceedings of IJCNN 2018}

\begin{document}
%

\author{\IEEEauthorblockN{Tinish Bhattacharya, Vivek Parmar and Manan Suri}
\IEEEauthorblockA{Department of Electrical Engineering, Indian Institute of Technology-Delhi, Hauz Khas, New Delhi - 110016\\
Corresponding Author: manansuri@ee.iitd.ac.in}
}

%


\maketitle
\copyrightnotice

\begin{abstract}
In this paper, we present \textit{MASTISK} (MAchine-learning and Synaptic-plasticity Technology Integrated Simulation frameworK). MASTISK is an open-source versatile and flexible tool developed in MATLAB for design exploration of dedicated neuromorphic hardware using nanodevices and hybrid CMOS-nanodevice circuits. 
MASTISK has a hierarchical organization capturing details at the level of devices, circuits (i.e. neurons or activation functions, synapses or weights) and architectures (i.e. topology, learning-rules, algorithms). In the current version, MASTISK provides user-friendly interface for design and simulation of spiking neural networks (SNN) powered by spatio-temporal learning rules such as Spike-Timing Dependent Plasticity (STDP). Users may provide network definition as a simple input parameter file and the framework is capable of performing automated learning/inference simulations.
Validation case-studies of the proposed open source simulator will be published in the proceedings of IJCNN 2018. The proposed framework offers new functionalities, compared to similar simulation tools in literature, such as: (i) arbitrary synaptic circuit modeling capability with both identical and non-identical stimuli, (ii) arbitrary spike modeling, and (iii) nanodevice based neuron emulation. The code of MASTISK is available on request at: \url{https://gitlab.com/NVM_IITD_Research/MASTISK/wikis/home}. 
\end{abstract}


%
\IEEEpeerreviewmaketitle

\section{Introduction}
\label{simu1}

Pure CMOS hardware based Spiking Neural Networks (SNN) designs have turned out to be less efficient owing to the large computation/memory requirement \cite{jo2010nanoscale}. Several hybrid CMOS-nanodevice designs involving emerging non-volatile memory nanodevices have been proposed as efficient SNN hardware alternatives compared to pure CMOS design. The emerging non-volatile memory devices offer added efficiency as they are ultra compact, low-power and intrinsically mimic properties of biological neurons and synapses \cite{burr2017neuromorphic}. Design exploration of such dedicated hybrid SNN hardware (CMOS-nanodevice) becomes a very complex and challenging multi-dimensional problem due to following reasons:
(i) need to capture different variants of spatio-temporal learning rules, neuron and synaptic behavior models \cite{andrew2003spiking,ghosh2009spiking,gerstner2002spiking}, (ii) wide variety of nanodevice/nanocircuit options for emulating neuro-synaptic characteristics with different underlying physics of operation, (iii)`lack of maturity at the level of device fabrication, (iv) almost negligible standardization when it comes to choice of synapse, neuron and learning-rule models, (v) wide-variety of application tasks, and (vi) lack of standard benchmarking. 
Comprehensive simulation frameworks or CAD tools which allow flexibility at the level of hardware-algorithm co-optimization thus become a necessity for addressing a domain like dedicated SNN hardware. Evaluating the impact of emerging device technology on network performance and associated hardware overheads by using a high-level simulation framework which takes in account details at the levels of device, circuit and algorithm simultaneously is essential to optimize performance and cut down on expensive ASIC manufacturing cycles. In literature, different biological SNN simulator frameworks have been reported such as: SpikeNET \cite{delorme2003spikenet}, Brian \cite{goodman2008brian}, Nengo \cite{bekolay2014nengo}, CARLsim3 \cite{beyeler2015carlsim}, NEST \cite{gewaltig2007nest}, NEURON \cite{carnevale2006neuron} etc. A limited number of simulation frameworks that deal with the implementation details of nanodevices in neural networks have also been reported, for instance: Xnet \cite{bichler2013design} and N2D2\cite{N2D2} are C++ based event-driven simulators, N2S3\cite{boulet2017n2s3} is a Scala-based event driven simulator, NeuroSim+ \cite{cpy17} is a device-to-algorithm framework for evaluating performances of emerging memory devices as synapses in Multi-Layer Perceptron (MLP) architectures. Most of these simulators can be either classified as Clock-Driven (CD) or Event-Driven (ED) based on when they update variables (at  all time-steps or at certain event triggered time-steps). In this work we present a new simulation tool developed in MATLAB called MASTISK. MASTISK is an acronym for \textit{MAchine-Learning and Synaptic-plasticity Technology Integration Simulation frameworK}. According to etymology of the word MASTISK (pronounced as \textit{mas-tea-she-q}), it means 'brain' in Sanskrit. Key distinguishing features of the current version of MASTISK, compared to other neuromorphic simulators reported in literature \cite{bichler2013design,N2D2,boulet2017n2s3}, are: 
\begin{itemize}
\item{It can model a wide variety of synaptic circuits, including those in which non-identical pulses are required for conductance modulation.}
\item{It allows users to emulate LIF/IF neuron circuits using nanodevices.}
\item{It allows emulation of any arbitrarily shaped neuron spike thus leading to different flavors of STDP or other spatio-temporal learning rules.}
\item{The tool has a parametric file interface, which requires the user to only specify a set of parameters defining the network topology, without the need for writing any additional code or script.}
\item{The tool provides an automated parameter tuning framework based on genetic algorithms to assist users in optimizing the network performance.}
\end{itemize}

\section{The MASTISK Framework}
\label{mastisk}

MASTISK is designed using a CD simulation approach. Following sub-sections describe the simulator- workflow, dataflow and cases. 


\begin{figure}[!ht]
\centering
\includegraphics[scale=0.38]{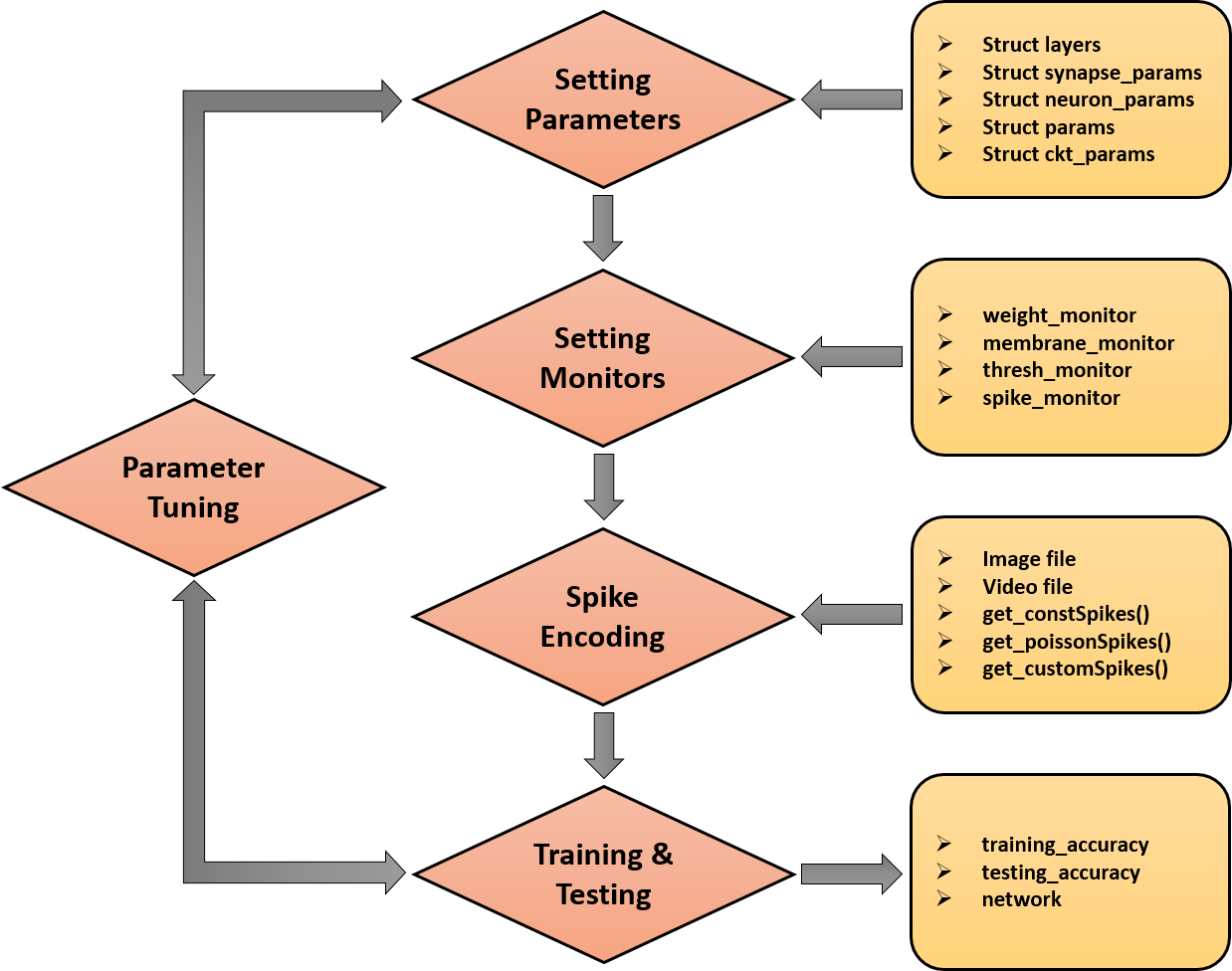}
\caption{Generic workflow of MASTISK}
\label{workflow}
\end{figure}

\begin{figure}[!t]
\centering
\includegraphics[scale=0.3]{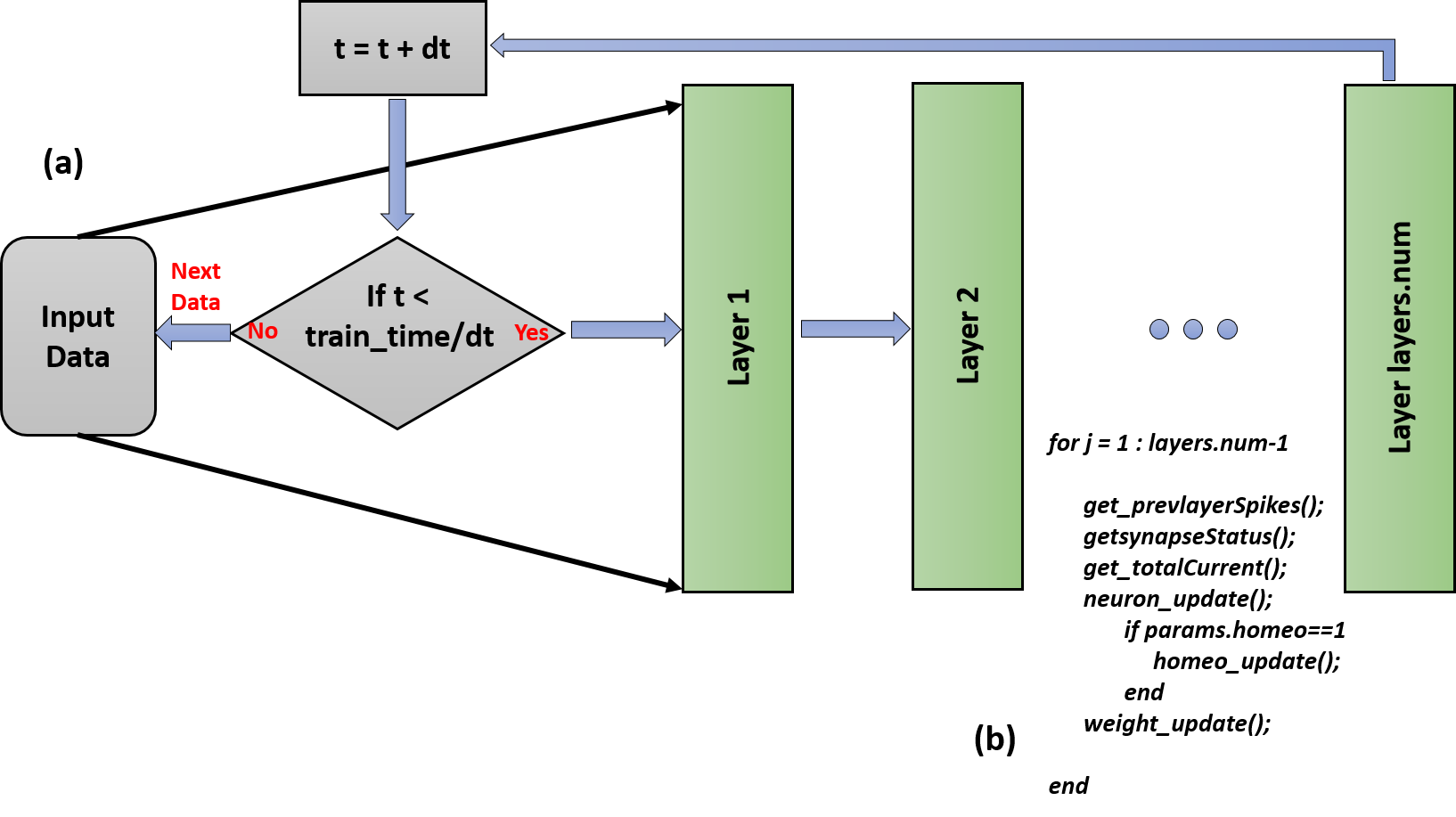}
\caption{(a) Flow of data and spikes along a generic network initiated by MASTISK. (b) Code overview of the basic functioning of the framework.}
\label{network1}
\end{figure}

\subsection{Simulation Workflow}
\label{simu2}
Key sequence for running any simulation in the proposed MASTISK framework is shown in Fig. \ref{workflow}. First step requires a set of override parameters from the user corresponding to the devices and circuits being used as building blocks as well as the network parameters. The parameters are stored in the properties of different structures as shown in Fig. \ref{workflow}. In SNNs the input data is encoded into multiple spike trains and the distribution of their frequencies determine the type of encoding used. MASTISK offers features for poisson, fixed frequency and custom AER sensor data based spike encoding. Following this, training and inference takes place which provides \textit{training\_accuracy}, \textit{testing\_accuracy} and the structure \textit{net} containing the network topology and weights of synapses, as output. The parameter tuning interface is optional and can be used to determine the set of parameters that leads to optimum network performance. 

\subsection{Simulator Dataflow}
The logical flow of information in the network and the simulator backbone is shown in Fig. \ref{network1}. After spike encoding of input is completed the output is computed for each layer sequentially beginning with layer 1 for the number of time steps N given by Eq. \ref{tstep},

\begin{equation}
\label{tstep}
Total\ time\ steps\ (N) = \dfrac{Total\ training\ time\ (T)}{Simulation\ time\ step\ (dt)}
\end{equation}

where dt is the simulator time-step and T is the total time for which the network needs to be trained. The function \textit{get\_prevlayerSpikes()} is used to obtain the spike information of the previous layer. The function \textit{getsynapsestate()} is used to obtain the modes (explained in Section \ref{simu3}) of the synapses and \text{get\_totalCurrent()} is used to obtain the total current entering each neuron. The neuron membrane potential and weights of synapses are updated using \textit{neuron\_update()} and \textit{weight\_update()} respectively.

\subsection{Simulation Cases}
\begin{figure}[!t]
\centering
\includegraphics[scale=0.35]{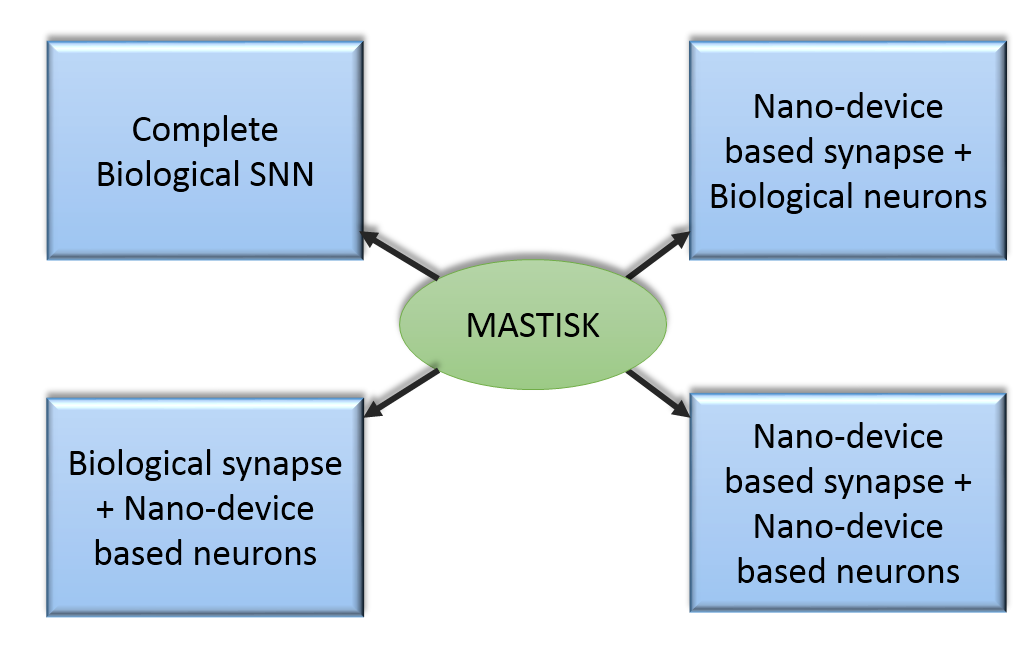}
\caption{The different facets of MASTISK.}
\label{edisonn}
\end{figure}

The framework in its current form offers four different simulation scenarios: (1) Biological SNN, (2) nanodevice based synapses, (3) nanodevice based LIF neurons, and (4) nanodevice based synapse and neurons, (Fig. \ref{edisonn}). Case (II) is for devices that show controlled conductance modulation properties like RRAM, PCM etc, in response to applied signal in the form of voltage or current. Case (III) is for devices that mimic the Leaky-Integrate-Fire (LIF) properties of membrane potential in a biological neuron. Few proposals incorporate all-nanodevice based implementation of SNN (neuronal and synaptic) \cite{sengupta2016vision,pantazi2016all}, represented by case (IV).

\section{Framework Capabilities}
\label{frame}
\begin{figure}[!b]
\centering
\includegraphics[width=\columnwidth]{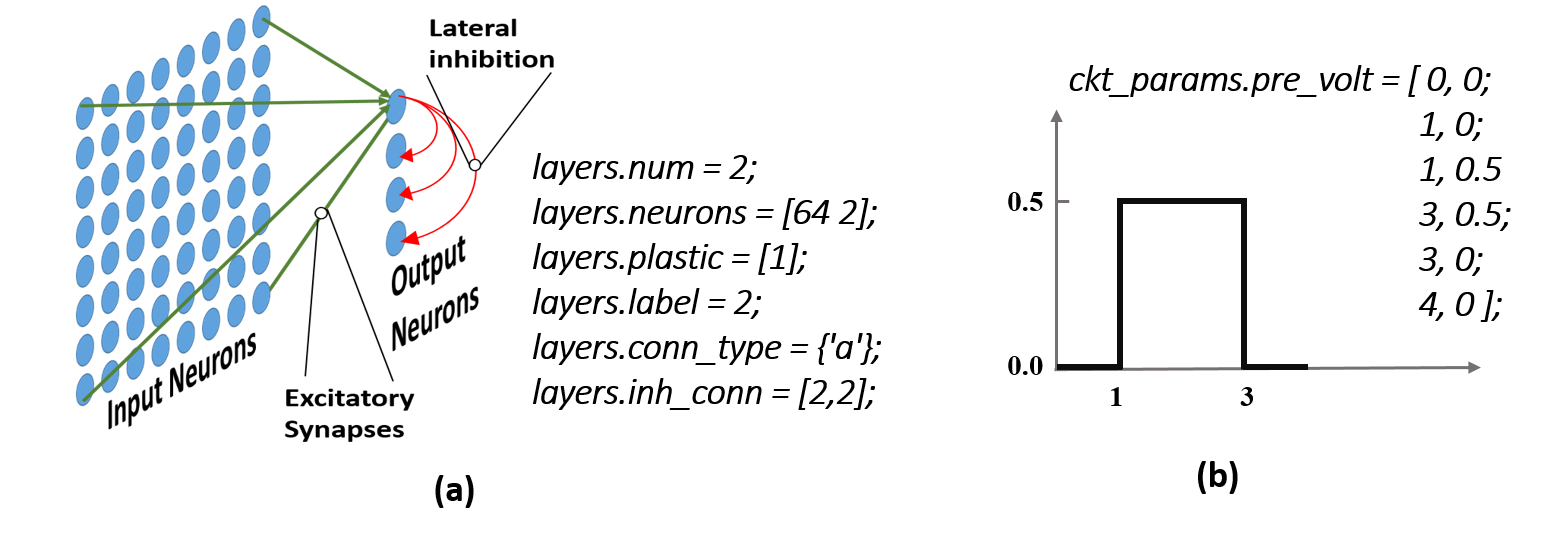}
\caption{ Example SNN with (a) Two layers and intra-layer inhibitory connections, (b) Three layers and inter-layer inhibitory connectons.}
\label{network2}
\end{figure}

\subsection{Network Topology Modeling}
The structure \textit{layers} is used to define the network topology. 
One of the most commonly used feed-forward SNN topologies in literature is shown in Fig.\ref{network2} (a). The number of layers and number of neurons in each layer are set by the \textit{num} and \textit{neurons} properties. MASTISK offers the flexibility to keep a certain set of weights fixed using \textit{layers.plastic} (Fig. \ref{network2} (a)), which in some cases can lead to improved performance and faster convergence in learning \cite{bengio2007greedy}. The output labeling layer in SNNs can be set using \textit{layers.label}. Different types of synaptic connections supported are: all-to-all, one-to-one and sparse, specified by \textit{layers.conn\_type} (Fig. \ref{network2} (a)). In case of sparse connectivity the degree of sparseness is specified by the value of \textit{layers.sparse}. Indices of start and end layers between which inhibitory connections exist are stored in \textit{layers.inh\_conn}.

\subsection{Spike Modeling}
The capability to model arbitrary spike shapes is highly desirable in a neuromorphic simulator. MASTISK applies piece-wise linear approximation to implement spike modeling. The points of non-differentiability in the spike waveform (both timestamp and amplitude) are stored in an array: \textit{ckt\_params.pre\_volt}. The simulator divides a complex spike into several pieces by using the specified starting (t$_{i}$,V$_{i}$) and end points (t$_{i+1}$,V$_{i+1}$) of each piece. If t$_{i}$ $\neq$ t$_{i+1}$, then points V$_{i}$ and V$_{i+1}$ are joined by a line as shown in Fig. \ref{network2} (b). If t$_{i}$ = t$_{i+1}$ and V$_{i}$ $\neq$ V$_{i+1}$, then there is a discontinuity or abrupt change between V$_{i}$ and V$_{i+1}$. In this case, V$_{i+1}$ becomes the starting point for the pieces following t$_{i+1}$.

\begin{figure}[!t]
\centering
\includegraphics[scale=0.38]{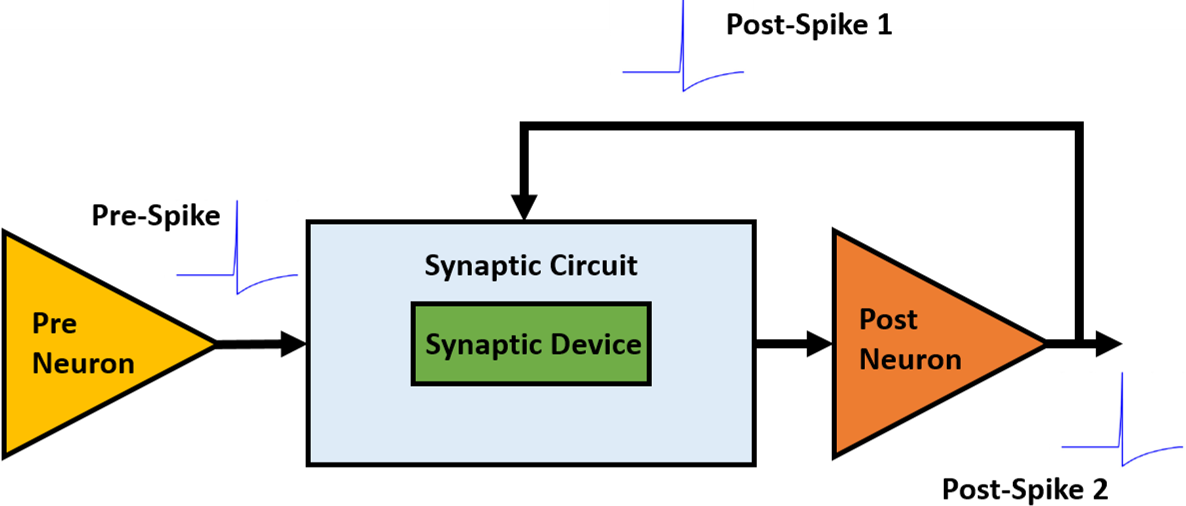}
\caption{Generic nanodevice based synaptic circuit implemented in MASTISK.}
\label{general_synapse}
\end{figure}

\subsection{Synaptic device and architecture modeling}
\label{simu3}
Generic architecture of a synaptic circuit that can be implemented in MASTISK is shown in Fig. \ref{general_synapse}. MASTISK considers three different types of neuronal spikes: post-spike1 (propagated back to synaptic circuit), post-spike2 (pre-spike for the next layer of synapses) and Inhib-spike (optional and used only when inhibition is to be implemented). Synaptic circuit and applied spikes depend on the synaptic device being used and this in turn effects the learning rule being implemented. Modeling such device specific constraints would require extensive code modifications in available simulation frameworks \cite{bichler2013design,N2D2,boulet2017n2s3}. In order to resolve this issue, we have split the operating states of a synapse in four modes: (1) Idle state, (2) Spike Transmission, (3) Potentiation and (4) Depression. In idle state there is neither any current flow through the synaptic device nor any conductance change. In spike-transmission state only current flows through the synapse and no conductance change takes place. Potentiation/Depression states implement increase/decrease of synaptic conductances.
\begin{figure}
\centering
\includegraphics[width=\columnwidth]{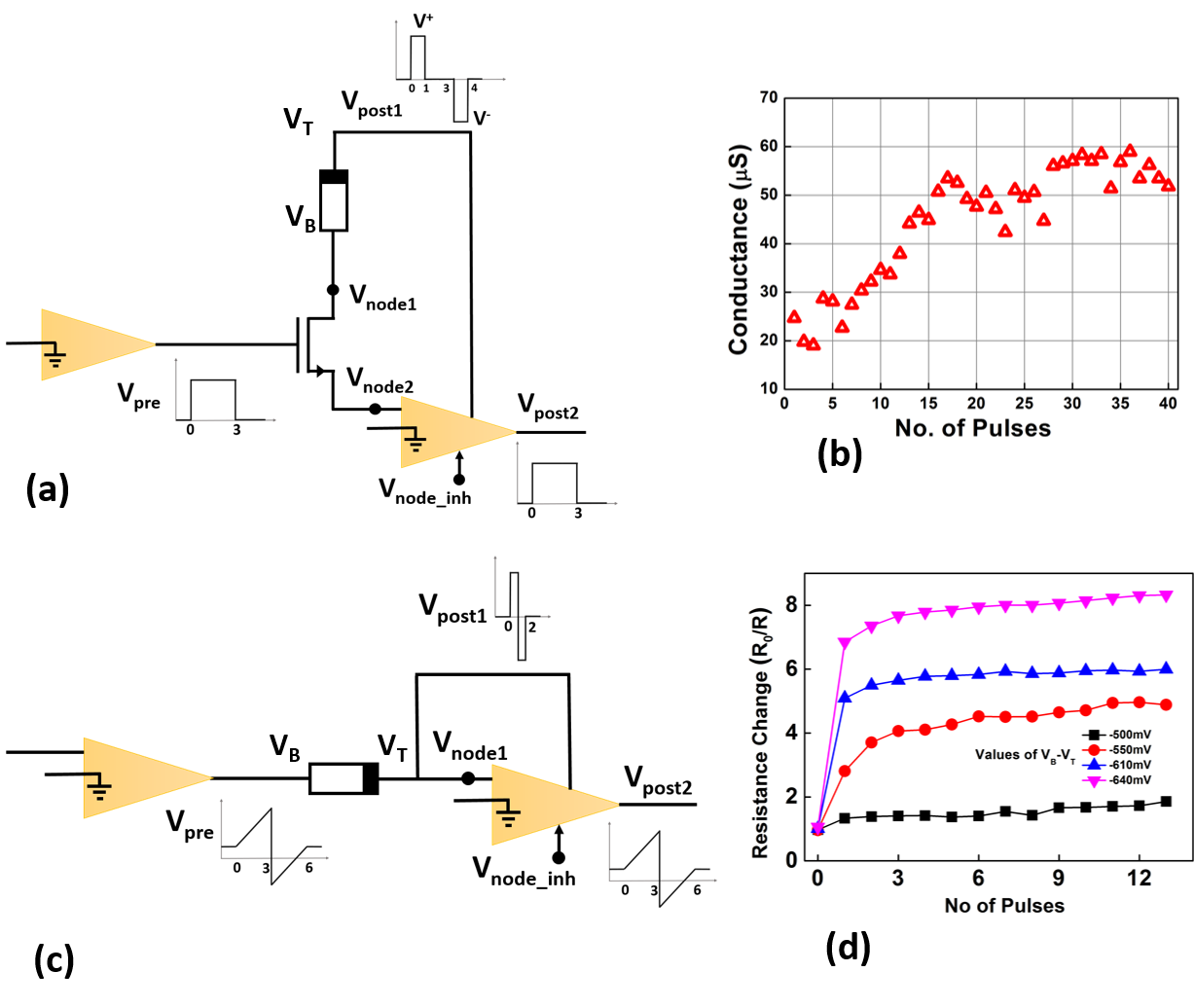}
\caption{MASTISK based implementations of: (a) 1T-1R synaptic circuit and spike scheme, adopted from \cite{ambrogio2016neuromorphic}(Inset (I) shows corresponding STDP characteristics for different initial synaptic resistance values.) (b) synaptic circuit and spike scheme adopted from \cite{covi2016analog} (Inset (I) shows resultant voltage across synaptic device (V$_{B}$-V$_{T}$) for different temporal orientation of pre/post-spikes, Inset (II) shows corresponding STDP characteristics for different initial synaptic resistance values.) (c) Inhibitory connection circuit. Synaptic conductance modulation in AlO$_{x}$/HfO$_{2}$ based RRAM device due to application of: (d) identical potentiating pulses, (e) identical depressing pulses, adopted from \cite{woo2016improved}. Synaptic conductance modulation in TiN/HfO$_{2}$/Ti/TiN based RRAM device due to application of (f) varying amplitude potentiating, and (g) varying amplitude depressing pulses, adopted from \cite{covi2016analog}.}
\label{master_synapse}
\end{figure}
The state of the synaptic circuit can be classified based on the presence or absence of the pre and post-neuronal spikes in any one of the following states: (1) no spikes, (2) only Pre-Spike, (3) only Post-Spike, and (4) both spikes. The functionality of any synaptic architecture can be implemented by establishing the mapping between the states of the synaptic device and the circuit. Two test cases of synaptic circuits and spikes adopted from \cite{ambrogio2016neuromorphic} and \cite{covi2016analog} along with their implementation in MASTISK are shown in Fig.\ref{master_synapse}(a) and (c). For emulation of STDP or any other spatio-temporal rule the exact timing of potentiation and depression relative to spike timing is required. The spike timings and synaptic states are taken care of different parameters. For details on the usage of these parameters, readers are referred to \cite{mastisk}. The circuit shown in Fig. \ref{master_synapse} (a) undergoes spike transmission only when there is a pre-spike (V$_{pre}$) (V$_{TB}$ = V$_{post1}$-V$_{node1}$, where V$_{post1}$ is a fixed DC bias when there is no post-spike), potentiation when the portion marked V$^{+}$ in V$_{post1}$ coincides with V$_{pre}$ (V$_{TB}$ = V$^{+}$-V$_{node1}$ \textgreater V$_{th}$) and depression when V$^{-}$ coincides with V$_{pre}$ (V$_{TB}$ = V$^{-}$-V$_{node1}$ \textless -V$_{th}$). The equation for current through device during spike transmission is captured by \textit{ckt\_params.ex\_eqs} and can be defined by the user. For simplicity we have assumed V$_{node1}$ = V$_{node2}$, when MOSFET is turned ON during pre-spike. However for more accurate circuit modeling one can define the relation between V$_{node1}$ and V$_{node2}$ in \textit{ckt\_params.ex\_eqs}. The device behavior during potentiation and depression are captured by obtaining experimental conductance modulation data of the device as shown in Fig. \ref{master_synapse} (b). The conductance states are stored in the arrays: \textit{device\_params.synapse\_levels\_ltp} and \textit{device\_params.synapse\_levels\_ltd}. However the conductance modulation characteristics shown in Fig. \ref{master_synapse} (b) are only good for synaptic circuits where at any point of time, conductance change is caused by identical pulses as in Fig. \ref{master_synapse} (a). In the circuit shown in Fig. \ref{master_synapse} (c), the resultant voltage across the device (V$_{TB}$ = V$_{post1}$ - V$_{pre}$) leads to conductance change when it crosses a threshold (V$_{TH}$). The magnitude of V$_{TB}$ that crosses (V$_{TH}$) keeps changing with the temporal orientation of the spikes w.r.t each other. In such case, experimental data showing conductance variation with number of applied pulses for different pulse widths or amplitudes is required (Fig. \ref{master_synapse} (d)). The expression for resultant voltage (V$_{TB}$) can be set by user using \textit{ckt\_params.v\_app}, and the experimental conductance values can be accommodated in \textit{ckt\_params.v\_example} in form of a 2D matrix, where each row signifies a different pulse amplitude or width.

\begin{figure}[!t]
\centering
\includegraphics[scale=0.38]{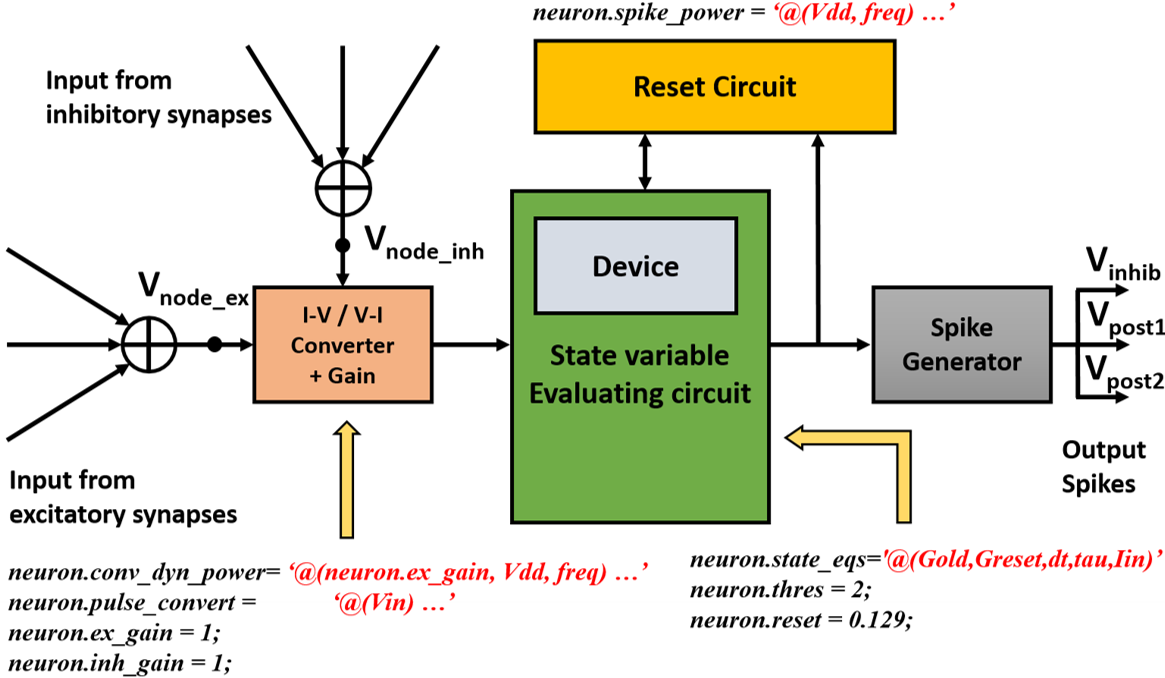}
\caption{Generic nanodevice neuron architecture implemented in MASTISK.}
\label{neuron}
\end{figure}

\begin{figure}[!t]
\centering
\includegraphics[scale=0.20]{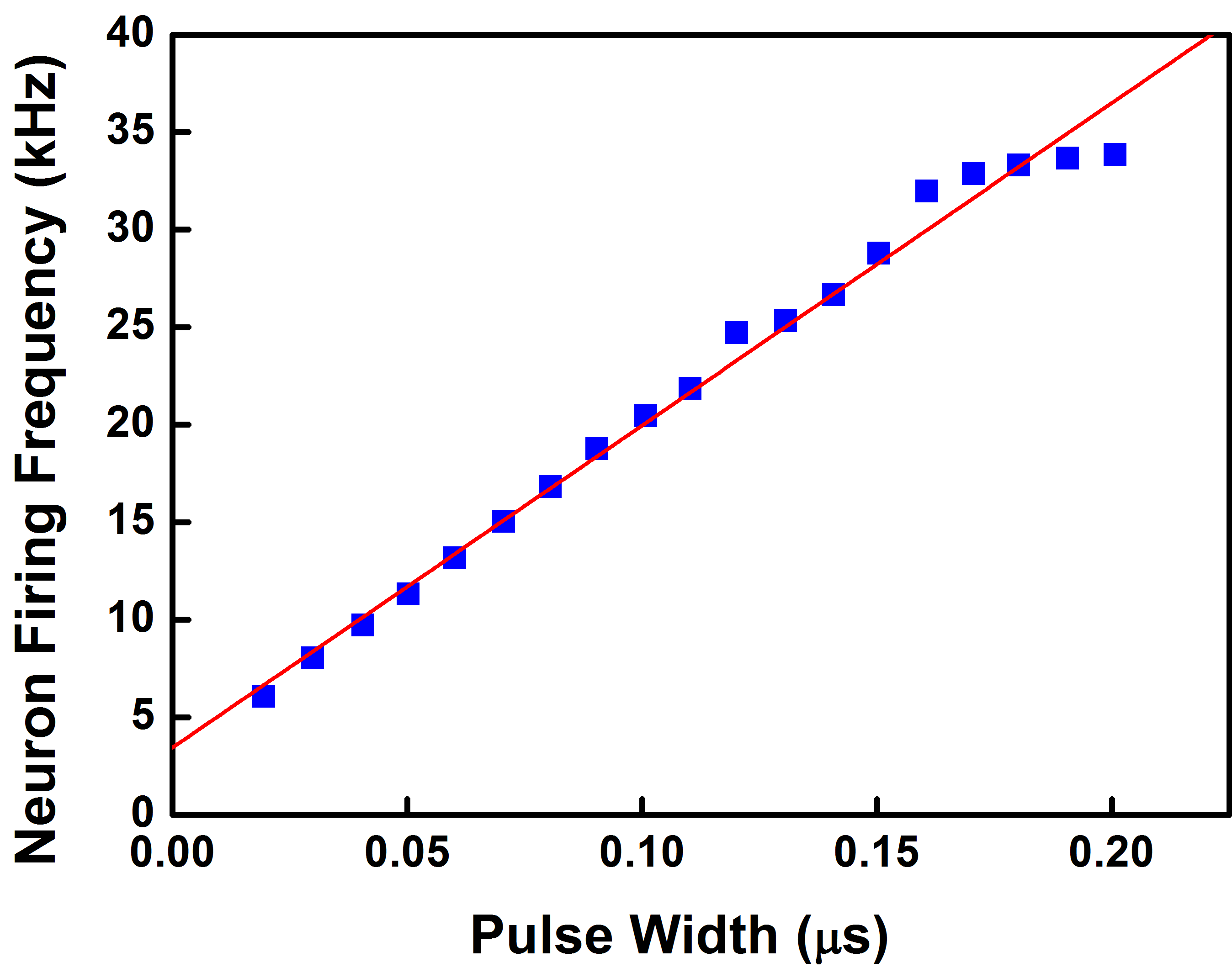}
\caption{Neuron frequency vs input pulse width adopted from \cite{tuma2016stochastic}.}
\label{firing_freq}
\end{figure}

\subsection{Neuron device architecture modeling}
Numerous devices have been proposed for implementing LIF or IF neurons in SNNs \cite{dutta2017leaky,tuma2016stochastic,sengupta2016magnetic,jaiswal2017proposal}. These devices have a state-variable that gets integrated with incoming signals. In order to simulate neuron functionalities with these devices, generalized architecture of neuron circuit implemented in MASTISK is shown in Fig. \ref{neuron}. V$_{node\_ex}$ and V$_{node\_inh}$ denote the voltages of the nodes where the summed signals from excitatory and inhibitory synapses enter respectively. The signal is then passed through a current-voltage/voltage-current converter or a buffer circuit. The user can specify the equation that best describes the dynamic power dissipated in their circuit with \textit{neuron\_conv\_dyn\_power}. The incoming signals may need to be converted to either fixed-amplitude, variable-width signal or vice versa. This width-amplitude characteristic can be defined using the function \textit{neuron.pulse\_convert()}. Neuron modeling in MASTISK is achieved by taking the neuron firing frequency variation with the strength of integrating signal as an input. Fig. \ref{firing_freq} shows frequency modulation with applied pulse width of PCM based IF neuron adopted from \cite{tuma2016stochastic}. This data can be used to extract the time constant (\textit{neuron.tau}) and threshold of the neuron device (\textit{neuron.thres}). LIF or IF equation governing the neuron state-variable updation is set by \textit{neuron.state\_eqs}. The state variable evaluating circuit (generally a voltage divider) generates output which is proportional to the conductance of the device and is evaluated against a threshold. Once the neuron spikes the state-variable is reset by the reset circuit.

\section{Discussion and Limitations}
\label{limit}
The accuracy of simulation results depends heavily on the choice of simulator time step (dt). Glitches or noisy spikes can only be observed if the time step of the simulation is smaller in comparison to the time period of the fastest spike. Present version of the proposed simulator does not incorporate direct integration with circuit simulators and as a result parameters such as thermal noise in devices and delays in analog circuits are not taken into consideration. However the modular structure of the simulator makes it possible for interfacing with circuit simulators in future. For example, functions governing spike transmission, dynamic power consumption etc, can be provided by user in form of equations or can be defined by a SPICE circuit simulator. Exporting the entire framework to Python, providing a GUI for input parameters, GPU acceleration to improve simulation time are some of the proposed improvements for future version of the present framework. We also would like to further extend the framework's capability to support other learning algorithms such as MLP, CNN, ELM \cite{suri2015oxram} and RBM \cite{suri2015neuromorphic,parmar2018design} in inference mode. 

\section{Conclusion}
\label{conc}
In this paper we present a new MATLAB based integrated technology benchmarking clock-driven SNN simulator framework called MASTISK. We describe our simulator's different functionalities like network modeling, spike modeling, unique synaptic circuit modeling strategy with both identical and non-identical conductance modulation pulses and parameter tuning using genetic algorithm. Unlike the rest, our simulator can also implement nanodevices along with its appropriate circuitry as neurons in SNNs. We show how the different parameters can be set to achieve the above functionalities and also present two case studies in which a RRAM based synapse and a PCM based neuron based SNN is simulated with actual experimental data and circuit topology extracted from literature. Further detailed extension of this work, including multiple device case-studies, parameter optimization etc. will be published in the proceedings of IJCNN 2018.


\section*{Acknowledgment}
This research activity under the PI Prof. M. Suri is partially supported by the Department of Science \& Technology (DST), SERB-EMR Government of India and IIT-D FIRP grants.



%
\bibliographystyle{IEEEtran}
\bibliography{main}

\end{document}